\documentstyle[12pt]{article}
\newcommand{\slp}{\raise.15ex\hbox{$/$}\kern-.57em\hbox{$\partial$}}
\newcommand{\sla}{\raise.15ex\hbox{$/$}\kern-.57em\hbox{$a$}}
\newcommand{\slA}{\raise.15ex\hbox{$/$}\kern-.57em\hbox{$A$}}
\newcommand{\slb}{\raise.15ex\hbox{$/$}\kern-.57em\hbox{$b$}}
\newcommand{\be}{\begin{equation}}

\newcommand{\ee}{\end{equation}}
\newcommand{\bear}{\begin{eqnarray}}
\newcommand{\ear}{\end{eqnarray}}
\newcommand{\eear}{\end{eqnarray}}
\newcommand{\ba}{\begin{eqnarray*}}
\newcommand{\ea}{\end{eqnarray*}}
\begin{document}
\begin{titlepage}
\setcounter{page}{1}
\begin{flushright}
HD--THEP--02--24\\
\end{flushright}
\vskip1.5cm
\begin{center}
{\large{\bf Lagrange versus Symplectic Algorithm for Constrained Systems}}\\
\vspace{1cm}
Heinz J. Rothe \footnote{email:  h.rothe@thphys.uni-heidelberg.de}
and Klaus D. Rothe \footnote{email:                      k.rothe@thphys.uni-heidelberg.de}
\\
{\it Institut  f\"ur Theoretische Physik - Universit\"at Heidelberg}
\\
{\it Philosophenweg 16, D-69120 Heidelberg, Germany}

{(August 2002)}
\end{center}

\begin{abstract}
\noindent
The systematization of the purely Lagrangean approach to constrained systems
in the form of an algorithm involves the iterative construction of a generalized
Hessian matrix $W$ taking a rectangular form. This Hessian will exhibit as many left zero-modes as there are
Lagrangean constraints in the theory. We apply this approach to a
general  Lagrangean in the first order formulation and show how
the seemingly overdetermined set of equations is solved for the velocities
by suitably extending $W$ to a rectangular matrix. As a
byproduct we thereby
demonstrate the equivalence of the Lagrangean approach to the traditional
Dirac-approach. By making use of this equivalence we show that a
recently proposed 
symplectic algorithm  does not necessarily reproduce
the full constraint structure of the traditional Dirac algorithm. 

\end{abstract}
\end{titlepage}

\newpage

\section{Introduction}
 
A number of algorithms have been developed over the past years for treating constrained Hamiltonian systems. Perhaps the most familiar one to the physicist
community is the one developed by Dirac \cite{Dirac64}.  Although very
elegant and powerful in its algebraic structure, this algorithm has been critizized for being based on the existence of so called ``primary
constraints", which are a purely phase-space artefact, and have no counterpart
on the Lagrangean level.
Faddeev and Jackiw \cite{Faddeev} have thus proposed an alternative method based
on a first order Lagrangean (symplectic) formulation, avoiding the introduction
of primariy constraints.
Furthermore, the local symmetries of the
Hamiltonian, as generated by the so called ``first-class" constraints
in Dirac`s terminology, turn out to be larger than those of the Lagrangean
This has led to a renewed interest in the problem of deducing the local symmetries of a Lagrangean from the
Hamiltonian formalism \cite{BRR}, and in particular 
to a revival of the ``Lagrangean approach", and the ``symplectic
approach"  to constrained systems
\cite{wozneto,Banerjee-neto,kimjkps1,kimjkps2,Proca,HKPR}.
Of all three methods, the Lagrangean algorithm is actually the most
pedestrian one, with a solid mathematical basis. The symplectic algorithm,
on the other hand, as developed in a series of elegant papers in refs.
\cite{wozneto}, lacks a rigorous mathematical
justification, and can lead, as we shall demonstrate, to an incomplete
solution of the problem. In order to establish the relation between the three
formalisms, we shall thus take as our starting point the Lagrangean 
approach as applied to first order Lagrangeans, in order to allow for a
comparison with the symplectic approach. As shown in section 2, the Lagrangean algorithm leads 
to a larger set of equations than the number of unknown velocities to be solved for. This is reflected in the fact that the generalized Hessian which
implements the algorithm, is a {\it rectangular} matrix
possessing as many left zero-modes, as there are Lagrangean constraints
hidden in the Euler-Lagrange equations. We show that these zero-modes are
of such a form, that they permit the solution of the equations of
motion in terms of the inverse of a quadratic matrix, whose elements 
are just the Poisson brackets of the Hamiltonian constraints - including
the primary constraints. We thereby establish the equivalence with
Dirac's algorithm. In section 3 we then consider the simple example
of the particle motion on a hypersphere and thereby demonstrate that the symplectic algorithm
of ref. \cite{wozneto}
is not always equivalent to the Dirac and Lagrangean approach. We conclude
this section by discussing the general condition under which this
symplectic algorithm fails. Section 5 summarizes our findings.

\section{The Lagrangean algorithm}

Given a second order Lagrangean, one can always find an equivalent first order Lagrangean of the form
\be\label{Lsymplectic}
L(Q,\dot Q) = a_{\alpha}(Q) \dot Q_\alpha - V(Q)
\ee
where $Q$ stands for $n$ degrees of freeedom $Q_\alpha\,,\alpha=1,2\cdot\cdot\cdot,n$.
The corresponding Euler-Lagrange equations read
\be\label{EL-equations}
W^{(0)}_{\alpha\beta}(Q) \dot Q_\beta = \frac{\partial V(Q)}{\partial Q_\alpha}\,.
\ee
where the matrix $W^{(0)}$ is defined by
\be\label{W-matrix}
W^{(0)}_{\alpha\beta}(Q) = \frac{\partial a_\beta}{\partial Q_\alpha}-
\frac{\partial a_\alpha}{\partial Q_\beta}
\ee

Let $r_0$ be the rank of the matrix $W^{(0)}$. Then there exist $n-r_0$ zero
modes of $W^{(0)}$, which we denote by $u^{(0)}(a)\,, 
a = 1 \,\cdot\cdot\cdot\, n-r_0$. Multiplying equations
(\ref{EL-equations}) from the left with these zero modes, we are led to 
the zero-level Lagrangean constraints
\be\label{zerolevel-constraints}
\varphi_a^{(0)} = \sum_{\alpha} u^{(0)}_\alpha(a) \frac{\partial V}{\partial Q_\alpha}=0\,,\quad a = 1,\cdot\cdot\cdot,n_0\,.
\ee
Some of these constraints may vanish identically. The remaining ones we
denote by $\varphi^{(0)}_{a_0}$. The corresponding zero modes $u^{(0)}(a_0)$ we refer to as ``non-trivial".

In general there are further constraints hidden in  equations 
(\ref{EL-equations}). In order to unravel them, we  implement 
their conservation by adjoining their time derivatives
\be\label{time-independence}
\left(\frac{\partial \varphi_{a_0}^{(0)}}
{\partial Q_\alpha}\right)\dot Q_\alpha = 0
\ee
to the equations (\ref{EL-equations}). This leads
to the following enlarged set of equations
\be\label{firstlevel-equations}
W^{(1)}_{A_1\beta}(Q)\dot Q_\beta = K^{(1)}_{A_1}(Q)
\ee
where $W^{(1)}_{A_1\beta}$ are now the elements of a rectangular
matrix 
\be
W^{(1)}_{A_1\beta}:=\left(
\begin{array}{c}
W^{(0)}_{\alpha\beta}\\
M^{(0)}_{{a_0}\beta}
\end{array}\right)\,\,,
\ee
with
\be\label{M}
M^{(0)}_{{a_0}\beta}= \frac{\partial \varphi_{a_0}^{(0)}}{\partial Q_\beta}
\ee
and
\be\label{K1}
K^{(1)}=\left(
\begin{array}{c}
\vec K^{(0)}
\\
\vec 0
\end{array}\right)\,.
\ee
where
\be
\vec K^{(0)} = \vec\nabla V(Q)
\ee 
We now look for ``non-trivial" zero modes ($u^{(1)}(a_1), a_1 = 1,\cdot\cdot\cdot,n_1)$
of $W^{(1)}$, and repeat the steps above, adjoining the time derivative
of any new constraints to  the equations of motion (\ref{firstlevel-equations}).
Repeating this algorithm, the iterative process terminates after
$L$ steps, when no new constraints are generated.

Denote the full set of constraints  
generated by the algorithm collectively by
$\{\varphi_{a}\}, a=1,\cdots,N$. Denote further the set $\{\alpha,a\}$ 
collectively by $\{A\}$. The {\it final} set of equations can then
be written in the form  
\be\label{eqnofmotion1}
W_{A\beta} \dot Q_\beta = K_{A}
\ee
where  
\be\label{rectangular}
W_{A\beta}:=\left(
\begin{array}{c}
W^{(0)}_{\alpha\beta}\\
M_{{a}\beta}\\
\end{array}\right)\,\,,
\ee
with
\be\label{Phialphabeta}
M_{a\beta}= \frac{\partial \varphi_a}{\partial Q_\beta}
\ee
and
\be\label{elth-K}
K_A:=\left(
\begin{array}{c}
\vec K^{(0)}
\\
\vec 0\\
\end{array}\right)\,.
\ee
Denoting by $\vec u(a)$ the left zero-modes of the matrix $W_{A\beta}$,
the constraints are given by $\varphi_a = \vec u(a)\cdot\vec K = 0$.
 
Equations (\ref{eqnofmotion1}) represent $n+N$ equations for the $n$ velocities
$\{\dot Q_\alpha\}$. In general such a set of equations would be overdetermined and
admit no non-trivial solution. Since the additional $N$ equations were however generated by a self-consistent algorithm from the original
Euler-Lagrange equations, the equations  (\ref{eqnofmotion1}) do in fact
admit a non-trivial solution. In the following we shall assume the first
order Lagrangean (\ref{Lsymplectic}) to describe a purely second class system in the Dirac terminology. In that case we have the following 

\bigskip\noindent
{\it Assertion:}

The unique solution  to (\ref{eqnofmotion1}) for the velocities is given by
\be\label{velocities}
\dot Q_\alpha = F^{-1}_{\alpha\beta}K^{(0)}_\beta
\ee
where $F^{-1}$ is the inverse of the matrix $F$ obtained by extending 
the rectangular matrix $W$ defined in (\ref{rectangular}) to the antisymmetric square matrix
\be\label{square} 
F_{AB}:=\left(
\begin{array}{cc}
W^{(0)}_{\alpha\beta}&-M^T_{\alpha b}\\
M_{{a}\beta}&{\bf 0}\\
\end{array}\right)\,\,,
\ee
with $M_{a \beta}$ defined in (\ref{Phialphabeta}), and
$M^T_{\alpha b} = M_{b\alpha}$.

\bigskip\noindent
{\it Proof of Assertion:}

Consider an enlarged space on which the square matrix (\ref{square}) is to
act (we streamline the notation in a self-evident way),
\be
\xi_A: = (Q_\alpha,\rho_a)
\ee
and the following  equations:
\be\label{eqnofmotion2}
F_{AB}\dot\xi_B = K_A
\ee
As we shall prove further below, $\det F \not = 0$ for a second class system.  Hence we can solve these equations for the velocities $\dot\xi_B$:
\be\label{auxiliary-velocities}
\dot\xi_A = F^{-1}_{AB}K_B\,.
\ee
We write the inverse matrix $F^{-1}$ in the form
\be\label{square-inverse} 
F^{-1}_{AB}:=\left(
\begin{array}{cc}
\tilde W^{(0)}_{\alpha\beta}&-\tilde M^T_{\alpha b}\\
\tilde M_{{a}\beta}&\omega_{ab}\\
\end{array}\right)\,\,.
\ee
Then $F^{-1}F = 1$ and $FF^{-1} = 1$ respectively imply, 
\bear\label{inverse-equations1}
&&\tilde W^{(0)}_{\alpha\gamma} W^{(0)}_{\gamma\beta} 
- \tilde M_{\alpha c}M_{c\beta} = \delta_{\alpha\beta}\nonumber\\ 
&&\tilde W^{(0)}_{\alpha\gamma} M^{T}_{\gamma b} = 0\nonumber\\
&&\tilde M_{a\gamma} W^{(0)}_{\gamma\beta} + \omega_{ac}M_{c\beta} = 0
\\
&&\tilde M_{a\gamma}M^{T}_{\gamma b} = -\delta_{ab}
\label{zeromode-normalization1}\,,
\ear
and
\bear\label{inverse-equations2}
&&W^{(0)}_{\alpha\gamma} \tilde W^{(0)}_{\gamma\beta} 
- M^T_{\alpha c}\tilde M_{c\beta} = \delta_{\alpha\beta}\nonumber\\ 
&&\tilde M_{a\gamma}\tilde W^{(0)}_{\gamma\beta} = 0\nonumber\\
&&W^{(0)}_{\alpha\gamma}\tilde M^T_{\gamma b} + M^T_{\alpha c}\omega_{cb} = 0\\
&&M_{a\gamma}\tilde M^{T}_{\gamma b} = -\delta_{ab}
\label{zeromode-normalization2}\,.
\ear
Consider eqs. (\ref{eqnofmotion2}) which, written out explictely, read
\be\label{EL-new}
W^{(0)}_{\alpha\beta}\dot Q_\beta - M^T_{\alpha b}\dot\rho_b = K^{(0)}_\alpha
\ee
\be\label{varphidot}
M_{a\beta}\dot Q_\beta = 0
\ee
From (\ref{Phialphabeta}) we see that the last equation states that $\dot\varphi_a = 0$, where $\varphi_a = 0$ are the 
constraints hidden in the equations of motion (\ref{EL-equations}). Because of this, requiring their presistance in time implies that the second term on the LHS of (\ref{EL-new}) must vanish. Making use of (\ref{zeromode-normalization1})  
this in turn implies that $\dot\rho_a = 0$ for all $a$. 
\footnote{An alternative  proof that this must indeed be the case will be given further below, where we make contact with the Hamiltonian 
formalism.} 
Setting $\dot\rho_a = 0$ in (\ref{auxiliary-velocities}), we have from
(\ref{square-inverse}), 
\be
\dot Q_\alpha = \tilde W^{(0)}_{\alpha\beta}K^{(0)}_\beta
\ee
\be\label{secondary-constraints}
0 = \tilde M_{a\beta}K^{(0)}_\beta
\ee

Eq. (\ref{secondary-constraints})  
is just the statement that $\varphi_a = 0$. To see this we notice that according to (\ref{inverse-equations1}), the vectors
\be
\vec u_A(a) := (\tilde M_{a\gamma},\omega_{ac})\
\ee
are just the left zero modes of the matrix (\ref{rectangular}). Hence
\be
\tilde M_{a\beta}K^{(0)}_\beta = u_A(a)K_A = \varphi_a \ .
\ee
As we now show, eqs. (\ref{auxiliary-velocities}) are nothing but the Hamilton equations of motion derived from the so-called extended Hamiltonian. By 
making contact with the Hamiltonian formalism, we will prove that i) $F$ is an invertible matrix, and ii) the solutions to 
(\ref{EL-new}) imply that $\dot\rho_a = 0$, as was claimed above. This, at the same time, will prove the uniqueness of the solution.

From the Dirac point of view, the symplectic Lagrangean (\ref{Lsymplectic}) descibes
a system with a {\it primary} constraint for every coordinate $Q_\alpha$:
\be\label{primaries}
\phi_\alpha := P_\alpha -a_\alpha(Q) = 0\,,\quad \alpha = 1,\cdots,n
\ee
where $P_\alpha$ are the momenta canonically conjugate to the
coordinates $Q_\alpha$. Since the  Lagrangean (\ref{Lsymplectic})
is first order in the time derivatives, the corresponding canonical Hamiltonian $H_c$ is just given by the potential $V$,
\be
H_c = V(Q)
\ee
and hence does not depend on the momenta. The dependence on the momenta
enters only in the {\it total} Hamiltonian via the primary
constraints:
\be
H_T(Q,P) = V(Q) + \sum_\alpha v_\alpha\phi_\alpha(Q,P) \,.
\ee
The Dirac algorithm will in general lead to secondary constraints, which we label by a latin index: $\varphi_a = 0$. It is easy to see that they are identical with the constraints generated by the Lagrangean algorithm. Thus
consider the persistance equations for the primary constraints
$\phi_\alpha$:
\[
\{\phi_\alpha,H_T\} = \{\phi_\alpha,V\} 
+ \sum_\beta \{\phi_\alpha,\phi_\beta\}v_\beta = 0 \ .
\]
From (\ref{primaries}) and (\ref{W-matrix}) we see that $\{\phi_\alpha,\phi_\beta\}= W^{(0)}_{\alpha\beta}(Q)$, so that the above
eqiations read,
\be\label{levelzeroEL}
W^{(0)}_{\alpha\beta} v_\beta = K^{(0)}_\alpha \ .
\ee
Multiplying this equation with the left-zero modes of $W^{(0)}$ we arrive 
at the level-zero Lagrangean constraints 
(\ref{zerolevel-constraints}), which are only functions of $Q$. Requiring their persistance in time 
as generated by $H_T$ yields $M^{(0)}_{a\beta} v_\beta = 0$, and adjoining
these equations to (\ref{levelzeroEL}),
\[
W^{(1)}_{A_1\beta}v_\beta = K^{(1)}_{A_1}\,.
\]
By taking appropriate linear combinations of these equations, new constraints 
may be generated which are functions of only the $Q_\alpha$'s. This just 
corresponds to looking for left-zero modes of $W^{(1)}$. The new constraints 
are thus identical with those derived in the Lagrangean approach at level ``one".
Proceeding in this way it is easy to see that the secondary constraints
generated by the Dirac algorithm applied to $H_T(Q,P)$ are
identical with the constraints $\{\varphi_a=0\}$ generated by the Lagrangean algorithm.

We now go over to the extended Hamiltonian by including the secondary constraints
with their respective Lagrange multipliers $\bar v_a$,
\be\label{H-extended}
H_T \to H_E = H_c + \sum_B\lambda_B\Omega_B \ ,
\ee
where
\[
\Omega_A: = (\phi_\alpha,\varphi_a)\,,\quad 
\lambda_A: = (v_\alpha,\bar v_a)\,.
\]
The Hamilton equations of motion for the coordinates $Q_\alpha$ associated with the extended Hamiltonian
$H_E$, read
\be\label{velocities}
\dot Q_\alpha = \{Q_\alpha,H_E\} = v_\alpha\,,
\ee
\be\label{momentum-dot}
\dot P_\alpha = \{P_\alpha,H_E\} = -K^{(0)}_\alpha + v_\beta\partial_\alpha a_\beta - \bar v_b\partial_\alpha\varphi_b
\ee
Consistency with the persistance in time of the primary constraints requires
\be
\dot\phi_\alpha = \dot P_\alpha - \dot a_\alpha = 0
\ee
One readily verifies from (\ref{momentum-dot}) that this leads to 
\be\label{EL-new1}
W^{(0)}_{\alpha\beta}v_\beta - M^T_{\alpha b}\bar v_b = K^{(0)}_\alpha \ .
\ee
On the other hand, persistance of the 
secondary constraints $\varphi_a$ leads to
\be\label{eqnsforv}
M_{b\beta}v_\beta = 0
\ee
Upon making use of (\ref{velocities}), we thus retrieved equations (\ref{EL-new}) and (\ref{varphidot}) if we identify $\bar v_a$ with $\dot\rho_a$. Hence in the Hamiltonian formalism these equations are merely the persistance equations of the primary 
and secondary constraints, which can be compactly written in the Hamiltonian form
\be\label{persistance-equations}
\{\Omega_A,H_c\} + \sum_B\{\Omega_A,\Omega_B\}\lambda_B = 0 \ .
\ee
We now recognize that the matrix elements of $F$ in (\ref{square}) are given by
\[
F_{AB} = \{\Omega_A,\Omega_B\} \ .
\]
Since the constraints have been assumed to be second class, this matrix is invertible. Noting further that
\[ 
\{\Omega_A,H_c\} =- K_A \ ,
\]
it follows from (\ref{persistance-equations}) that 
\be\label{lambda}
\lambda_A = F^{-1}_{AB}K_B \ .
\ee
With $\dot Q_\alpha = v_\alpha$, these equations are nothing but (\ref{auxiliary-velocities}), with 
$\dot\rho_a$ identified with $\bar v_a$.

To prove the equivalence of (\ref{EL-new1}) and (\ref{eqnsforv}) with the original set of equations (\ref{eqnofmotion1}), we must 
still show that eqs. (\ref{lambda}) imply that $\bar v_a = 0$. To this effect we recall that the secondary constraints $\varphi_a = 0$ have actually been generated 
by the {\it total} Hamiltonian $H_T$ from the persistance equations 
\[
\{\Omega_A,H_c\} + \sum_\beta \{\Omega_A,\phi_\beta\} \lambda_\beta = 0 \ .
\]
Consistency with (\ref{persistance-equations}) therefore requires that 
\[
\sum_b\{\Omega_A,\varphi_b\}\bar v_b = 0 \ ,
\]
or equivalently $M^T_{\alpha b}\bar v_b = 0$, which, upon making use of (\ref{zeromode-normalization1}), implies  
$\bar v_a = 0$. This completes the proof of our assertion.

Concluding this section we have therefore shown the full equivalence of the 
Lagrangean and Hamiltonian approach to the theory described by the first 
order Lagrangean (\ref{Lsymplectic}). Any other approach must therefore reproduce the constrained structure of the Lagrangean approach. An alternative (symplectic) algorithm for unravelling the constrained structure was proposed in ref. [4]. 
In the following section we will show that the symplectic algorithm does not necessarily generate the correct constrained structure.  

\section{The symplectic algorithm}

In the following we first illustrate in terms of a simple example, an alternative algorithm for generating the constraints, as proposed in ref. \cite{wozneto}. We refer to it as the ``symplectic algorithm".

\subsection{Particle on a Hypersphere}
 
The following (second order) Lagrangean is referred to as describing the non-linear sigma model in Quantum Mechanics:

\be
L = \frac{1}{2}\dot{\vec q}^2 + \lambda(\vec q^2 - 1)
\ee
where $\vec q = (q_1,\cdots,q_n)$. The equivalent symplectic Lagrangean reads,
\be
L^{(0)} = \vec p\cdot\vec{\dot q} - V^{(0)}
\ee
with 
\be
V^{(0)} = -\lambda(\vec q^2 - 1)+ \frac{1}{2}{\vec p}^2
\ee
The Lagrangean is of the form
\be\label{levelzero-Lagrangean}
L^{(0)} = a_\alpha(Q)\dot Q_\alpha - V^{(0)}(Q)
\ee
with
\be
Q_\alpha := (\vec q,\vec p,\lambda)\,,\quad
a_\alpha:= (\vec p,\vec 0,0) 
\ee
The equations of motion are of the form (\ref{EL-equations}), with
\be
W^{(0)} = \left(
\begin{array}{ccc}
{\bf 0}&-{\bf 1}&\vec 0\\
{\bf 1}&{\bf 0}&\vec 0\\
\vec 0^T&\vec 0^T&0
\end{array}
\right)
\ee
and 
\be
K^{(0)}_\alpha = \left(
\begin{array}{c}
-2\lambda\vec q\\
\vec p\\
-(\vec q^2 - 1)
\end{array}
\right)
\ee
The matrix $W^{(0)}$ has one ``zero-level" zero mode:
\[ u^{(0)}_\alpha := (\vec 0,\vec 0,1) \]
implying the constraint
\be\label{zerolevelconstraint}
\varphi^{(0)} = -u^{(0)}_\alpha K^{(0)}_\alpha = \vec q^2 - 1 =0
\ee
This constraint will necessarily coincide with that of the Lagrangean
approach at the zero'th level.

In the symplectic algorithm the time derivative
of the constraint (\ref{zerolevelconstraint}) is however added in the (partially integrated) 
form $-\dot\rho^{(0)}\varphi^{(0)}$ to the zero-level Lagrangean (\ref{levelzero-Lagrangean}),
\footnote{In ref. \cite{wozneto} the term proportional to the Lagrange multiplier $\lambda$ in $L^{(0)}$ has been absorbed into the
term proportional to $\dot\rho$.}
to yield the first level Lagrangean
\be
L^{(1)} = L^{(0)} -\dot\rho^{(0)}\varphi^{(0)}\,,
\ee
where $\rho^{(0)}$ is a new dynamical variable.
Correspondingly we define the extended set of coordinates
\[ \xi^{(1)}_{A_1}: = (\vec q,\vec p,\lambda,\rho^{(0)})\,. \]
$L^{(1)}$ can be written in the form
\[
L^{(1)} = a^{(1)}_{A_1}\dot\xi^{(1)}_{A_1} - V^{(0)}(Q)\,,
\]
where
\[ a^{(1)}_{A_1}: = (\vec p,\vec 0,0,-\varphi^{(0)}) \,.\]
The corresponding Euler-Lagrange equations read
\[
F^{(1)}_{A_1B_1}\dot\xi^{(1)}_{B_1} = K^{(1)}_{A_1}
\]
where the ``first-level" symplectic {\it square} matrix is given by,
\be
F^{(1)} = \left(
\begin{array}{cccc}
{\bf 0}&-{\bf 1}&\vec 0&-2\vec q\\
{\bf 1}&{\bf 0}&\vec 0&\vec 0\\
\vec 0^T&\vec 0^T&0&0\\
2\vec q^T&\vec 0^T&0&0
\end{array}
\right)
\ee
and
\be
K^{(1)} = \left(
\begin{array}{c}
-2\lambda\vec q\\
\vec p\\
-(\vec q^2 - 1)\\
0
\end{array}
\right)\,.
\ee
$F^{(1)}$ has two (level-one) zero modes, 
\[ 
u^{(1)}_{A_1}(1): = (\vec 0,\vec 0,1,0) 
\]
\[ 
u^{(1)}_{A_1}(2): = (\vec 0,2\vec q,0,-1) 
\]
The first zero mode reproduces the constraint $\vec q^2 - 1=0$.
The second zero mode yields the new constraint
\[
\varphi^{(1)} = \vec u^{(1)}(2)\cdot \vec K^{(1)} = 2\vec p\cdot \vec q
\]
As one easily verifies, these constraints are identical with those obtained in the Lagrangean algorithm described in section 1, at this level.

According to the symplectic algorithm we now define the second level Lagrangean
by adding the new constraint in the form
\[ 
L^{(2)} = L^{(0)} - \dot\rho^{(0)}\varphi^{(0)} - \dot\rho^{(1)}\varphi^{(1)}
\]
or
\[
L^{(2)}= a^{(2)}_{A_2}\dot\xi^{(2)}_{A_2} - V^{(0)}(Q)
\]
with
\[
\xi^{(2)}_{A_2}: = (\vec q,\vec p,\lambda,\rho^{(0)},\rho^{(1)})
\]
and 
\[ 
a^{(2)}_{A_2}: = (\vec p,\vec 0,0,-(\vec q^2 - 1),-2\vec p\cdot\vec q) 
\]
For the corresponding symplectic matrix one obtains
\be\label{F2}
F^{(2)} = \left(
\begin{array}{ccccc}
{\bf 0}&-{\bf 1}&\vec 0&-2\vec q&-2\vec p\\
{\bf 1}&{\bf 0}&\vec 0&\vec 0&-2\vec q\\
\vec 0^T&\vec 0^T&0&0&0\\
2\vec q^T&\vec 0^T&0&0&0\\
2\vec p^T&2\vec q^T&0&0&0
\end{array}
\right)
\ee
As one readily checks, this matrix has only one zero mode 
$u^{(2)}_{A_2}: = (\vec 0^T,\vec 0^T,1,0,0)$ 
which, however, just reproduces the constraint $\varphi^{(0)} = 0$.
Hence the symplectic algorithm terminates at this point, leaving
us with a non-invertible matrix.
\footnote{In ref. \cite{wozneto} the Lagrange multiplier $\lambda$ was absorbed into the dynamical variable $\rho^{(0)}$. Theirby the information about $\lambda$ was lost, and the resulting matrix $F^{(2)}$ at level 2 became invertible.}
On the other hand, one readily
checks that the standard Lagrangean (or equivalently, Dirac) algorithm generates
not only the constraints $\vec q^2 - 1=0, \, \vec p\cdot \vec q =0$,
but also one futher constraint $2\lambda \vec q^2 + \vec p^2 = 0$.
Indeed, in the Lagragian algorithm, $F^{(2)}$ in (\ref{F2}) is
replaced by the {\it rectangular} matrix
\be\label{W2}
W^{(2)} = \left(
\begin{array}{ccccc}
{\bf 0}&-{\bf 1}&\vec 0\\
{\bf 1}&{\bf 0}&\vec 0\\
\vec 0^T&\vec 0^T&0\\
2\vec q^T&\vec 0^T&0\\
2\vec p^T&2\vec q^T&0
\end{array}
\right)
\ee
which is seen to possess the three zero modes,
\be\label{level2-zeromodes}
u^{(2)}(1) = \left(
\begin{array}{c}
\vec 0\\
\vec 0\\
1\\
0\\
0
\end{array}
\right)`,,
\quad
u^{(2)}(2) = \left(
\begin{array}{c}
0\\
2\vec q\\
0\\
-1\\
0
\end{array}
\right)\,,
\quad
u^{(2)}(3) = \left(
\begin{array}{c}
2\vec q\\
-2\vec p\\
0\\
0\\
1
\end{array}
\right)\,,
\ee
which {\it in addition} to the constraints $\varphi^{(0)} = 0$, $\varphi^{(1)}=0$,
imply a new constraint
\be\label{constraint3}
\varphi^{(2)} := 2\lambda\vec q^2 + \vec p^2 = 0\,. 
\ee
Hence we are taken to a third level with the corresponding enlarged matrix given by
 \be
W^{(3)} = \left(
\begin{array}{ccc}
{\bf 0}&-{\bf 1}&\vec 0\\
{\bf 1}&{\bf 0}&\vec 0\\
\vec 0^T&\vec 0^T&0\\
2\vec q^T&\vec 0^T&0\\
2\vec p^T&2\vec q^T&0\\
4\lambda\vec q^T&2\vec p^T&2\vec q^2
\end{array}
\right)\,.
\ee
As one readily checks, $W^{(3)}$ has no {\it new} zero modes. Hence the algorithm terminates at this point. Notice that the extension of this matrix to a square matrix as discussed in section 2 results in an invertible matrix,
reflecting a second class system.

We see that the symplectic algorithm fails to generate the correct set
of constraints known to be present for the model in question. In fact,
from the point of view of the {\it second order} Lagrangean formulation
there exists just one primary constraint $\phi = p_\lambda =0$,
where $p_\lambda$ is the momentum conjugate to the variable $\lambda$,
and the total Hamiltonian correspondingly reads, 
$H_T = \frac{1}{2}\vec p^2 - \lambda (\vec q^2 - 1) + vp_\lambda$.
As one readily checks, the last constraint (\ref{constraint3})
just serves to fix the Lagrange multiplier $v$ in $H_T$ to $v=0$.
Only at this final stage the second class nature of the model in question
becomes evident. If we stop at level two, $v$ remains arbitrary,
as expressed by the zero column in (\ref{F2}) and (\ref{W2}).

\subsection{When does the symplectic algorithm fail?}

We now  examine in general terms at which point the symplectic algorithm
begins to fail. To this end we examine what the symplectic algorithm described above corresponds to on Hamiltonian level. Let $L^{(0)}$ be of the form
(\ref{Lsymplectic}), with (\ref{primaries}) the corresponding primary constraints.
At the $\ell+1$'th level, the symplectic algorithm  leads to a
Lagrangean of the form (we streamline the notation)
\[
L^{(\ell+1)} = L^{(0)} - \sum_{a_\ell}\dot\rho_{a_\ell}
\varphi_{a_\ell}(Q)\,,
\]
where $\varphi_{a_\ell}, a_\ell = 1,\cdots,n_\ell$ denote {\it all} the
constraints generated by the iterative procedure up to level $\ell$.
The corresponding total Hamiltonian reads,
\[
H^{(\ell+1)}_T = H^{(0)}_T + \sum_{a_\ell}
\lambda_{a_\ell}\phi_{a_\ell}\,.
\]
Here $\{\phi_{a_\ell}\}$  denote the corresponding set of primary constraints
associated with $\{\dot\rho_{a_\ell}\}$, 
\[
\phi_{a_\ell}=
P_{a_\ell} + \varphi_{a_\ell}(Q)\,,
\]
where $P_{a_\ell}$ are the momenta conjugate to the dynamical
variables $\rho_{a_\ell}$
and
\[
H^{(0)}_T = H^{(0)} + \sum_\alpha v_\alpha \phi_\alpha
\]
with $\phi_\alpha$ the primary constraints (\ref{primaries}), associated with the original Lagrangean $L^{(0)}$. Hence in the symplectic algorithm described above, the total Hamiltonian is modified at each level. Clearly the Euler-Lagrange equations
derived from $L^{(\ell +1)}$ and the Hamilton equations of motion 
following from  $H^{(\ell +1)}$ describe the same dynamics. 

Conservation of all the primary constraints requires,
\bear\label{timeindependence}
\{\phi_\alpha, H^{(\ell+1)}_T\}&=& -\frac{\partial V}{\partial Q_\alpha}
+ \sum_\beta \{\phi_\alpha,\phi_\beta\}v_\beta
+ \sum_{b_\ell} \{\phi_\alpha,\phi_{b_\ell}\} 
\lambda_{b_\ell} = 0\nonumber\\
\{\phi_{a_\ell}, H^{(\ell+1)}_T\} &=& 
 \sum_\beta \{\phi_{a_\ell},\phi_\beta\}v_\beta
+ \sum_{b_\ell} \{\phi_{a_\ell},\phi_{b_\ell}\} 
\lambda_{b_\ell} = 0.
\ear
Let $\Phi_{A_\ell}$ stand for 
\[
\Phi_{A_\ell} := (\phi_\alpha,\phi_{a_\ell}),
\]
with $\lambda_{A_\ell}$ the corresponding set of Lagrange multipliers:
\[
\lambda_{A_\ell} := (v_\alpha,\lambda_{a_\ell})
\]
Then we may write (\ref{timeindependence}) in the compact form 
\be\label{compact}
\sum_{B_\ell} \{\Omega_{A_\ell},\Omega_{B_\ell}\}
\lambda_{B_\ell} =  K_{A_\ell}\,,
\ee
where 
\[
\vec K=(\vec\nabla V,\vec 0)\,,
\]
with $\vec 0$  an $N_\ell= n+n_\ell$ - component null-vector.

One readily checks that  $\{\Omega_{A_\ell},\Omega_{B_\ell}\}$
is identical with $F_{A_\ell B_\ell}$ in (\ref{square}) at the $\ell$'th level.
Furthermore, with the identification of $v_\alpha$ and
$\lambda_{a_\ell}$ with $\dot Q_\alpha$ and $\dot\rho_{a_\ell}$
via the Hamilton equations of motion,
 \[
\dot Q_\alpha = \{Q_\alpha, H^{(\ell+1)}_T\} = v_\alpha\,,
\]
\[
\dot\rho_{a_\ell} = \{\rho_{a_\ell},H_T^{(\ell+1)}\}= \lambda_{a_\ell}
\]
we see that the {\it persistance} equations (\ref{timeindependence}) are just the equations of motion obtained from $L^{(\ell +1)}$ in the symplectic approach.

Within the Hamiltonian formalism, the search for zero modes  of $F$ at level $\ell$ now corresponds to
seeking  linear combinations of {\it all} the primaries, 
$ u_{A_\ell} \Phi_{A_\ell}$, such that 
\be\label{firstclass}
\sum_{A_\ell}u_{A_\ell} \{\Phi_{A_\ell},\Phi_{B_\ell}\}=0\,. 
\ee
From (\ref{compact}) we see that these equations imply linearly independent (non-trivial) constraints, which we denote by
\[
\varphi_{a_\ell}= \sum u_{A_\ell}(a_\ell)K_{A_\ell}=0\,.
\]
Of the conditions (\ref{firstclass}), only those with $B_\ell = \beta$,
\be\label{lagrange-constraints}
\sum_{A_\ell} u_{A_\ell}(a_\ell) \{\Phi_{A_\ell},\phi_{\beta}\}=0
\ee
are contained in the Lagrangean (and hence traditional Dirac) approach.
Let $u(a_\ell)$ be solutions of (\ref{lagrange-constraints}). From (\ref{firstclass}),
with $B_\ell = b_\ell$ we see that the symplectic approach thus implies 
the {\it additional} restrictions
\be\label{additional-restrictions}
u_{\alpha}(a_\ell) \frac{\partial\varphi_{b_\ell}}{\partial Q_\alpha}=0\,,
\ee
for the zero modes, which are {\it not} contained in the Lagrangean algorithm.
Hence we have a mismatch between the symplectic and Lagrangean algorithm, once the latter condition is not satisfied in the iterative process,
and the constraint structure becomes inequivalent for the two algorithms. This is the main point of this paper. 
\footnote{This subtle point has been missed in ref. \cite{Shirzad}}

Let us exemplify this for the case of the "particle on a hypersphere".
At the second level the Lagrangean algorithm
leads to the zero modes (\ref{level2-zeromodes}). We verify that at the
zero'th and first level of the iterative process the condition (\ref{additional-restrictions}) is still verified, whereas this is not the case for the second level zero mode $u^{(2)}(3)$ in (\ref{level2-zeromodes}), since
\[
u^{(2)}_\alpha(3) \frac{\partial\varphi^{(0)}}{\partial Q_\alpha}=4\vec q^2 
\not =0.
\]
This explains why the algorithm stops before generating one further
constraint, $\varphi^{(2)}=0$, eq. (\ref{constraint3}).

It is instructive to further ellucidate the meaning of this finding.
Going through the iterative procedure on Hamiltonian level (found above
to be equivalent to the symplectic algorithm), we arrive after the second
iterative step at the Hamiltonian
\ba
H^{(2)}_T &=& V^{(0)}(q,p,\lambda)+\sum_{i=1}^n (v_{q_i}(P_{q_i}-p_i)
+ v_{p_i}P_{p_i})
\nonumber\\
&+&  v p_\lambda + \lambda_1(P_1+\vec q^2 - 1) 
+ \lambda_2(P_2+2\vec q\cdot \vec p)
\ea
Conservation in time of the primaries now merely serves to fix all the
 Lagrange multipliers $\lambda, \lambda_1,\lambda_2,\lambda_2$, and leads to:
\ba
\vec v_{p} + 2\lambda_1\vec q + 2\lambda_2\vec p &=&2\lambda \vec q\\
\vec v_{q} + 2\lambda_2 \vec q &=& \vec p\\
\vec v_{q}\cdot \vec p + \vec v_{p} \cdot \vec q &=& 0\\
\vec v_{q}\cdot \vec q &=& 0\,,
\ea
as well as the constraint $\vec q^2 - 1=0$.
These equations may be solved for $\lambda_1$ and $\lambda_2$,
\ba
\lambda_1 &=& \frac{1}{2}(\vec p^2 + 2\lambda - 2\vec p\cdot \vec q)\nonumber\\
\lambda_2 &=& \frac{1}{2}(\vec p\cdot \vec q)\,,\nonumber
\ea
and hence vor $v_{q_i}$ and $v_{p_i}$, leaving $v$ undetermined.
Hence $\vec q^2 - 1=0$ is the only constraint (as reflected by the zero
column in (\ref{F2}) and (\ref{W2})), unless we set $\lambda_1 = \lambda_2 = 0$. This just corresponds to working with the total Hamiltonian $H^{(0)}_T$.
In that case
the algorithm does not terminate, but rather 
generates one further constraint, $\vec p^2 + 2\lambda=0$, whose time independence will finally fix also $v$ to vanish.

\section{Conclusion}

In this paper we have examined the interrelation between three
different algorithms currently in use for unravelling the constrained
structure of first order Langrangians. We have referred to these as
the ``Lagrangean", ``Dirac" and ``symplectic" algorithms.
Of these the first two rest on a solid foundation, and,
as we have seen, there exists a one-to-one correspondance between these 
formalisms. In particular we have shown how to invert the seemingly overdetermined system of equations of the Lagrangean algorithm. As for the symplectic algorithm presented in refs. 
\cite{wozneto}, it does not always reproduce the correct set of constraints, as we have seen. In fact, we have shown for a general first order Lagrangean, under what conditions the algorithm  fails to reproduce all of the constraints correctly. A concrete example has exemplified this.


\end{document}